# Surface-Assisted Luminescence: the PL Yellow Band and the EL of n-GaN Devices.

José Ignacio Izpura

Group of Microsystems and Electronic Materials. GMME-CEMDATIC
Universidad Politécnica de Madrid. 28040-Madrid. Spain.
e-mail: joseignacio.izpura@upm.es

## Abstract

Although everybody should know that measurements are never performed directly on materials but on devices, this is not generally true. Devices are physical systems able to exchange energy and thus subject to the laws of physics, which determine the information they provide. Hence, we should not overlook device effects in measurements as we do by assuming naively that photoluminescence (PL) is bulk emission free from surface effects. By replacing this unjustified assumption with a proper model for GaN surface devices, their yellow band PL becomes surface-assisted luminescence that allows for the prediction of the weak electroluminescence recently observed in n-GaN devices when holes are brought to their surfaces.

# 1. Introduction

Some years ago, we published reports about the high gain of GaAs photoresistors at low illumination levels due to the optical modulation of depleted regions under the GaAs surface [1, 2]. This gain $G_S$, derived from the photo-FET devices that the naked surface and the bottom interface with the substrate create in GaAs epilayers, is inversely proportional to the illumination power $P_L$ impinging on the photoresistor; thus, it is orders of magnitude greater than the "photoconductive" gain $G_B$ due to the conductivity modulation of the GaAs bulk by photogenerated electron-hole pairs at low illumination levels. Therefore, the changes $\Delta R$ in the resistance $R$ of any GaAs photoresistor under weak illumination power $P_L$ reflect modulations in the channel cross section of this two-terminal device (2TD) rather than conductivity modulations of their inner material as it is commonly assumed.

By defining the gain of these photoresistors by the ratio $G_X=\Delta R/\Delta P_L$, where $\Delta P_L$ is the change in optical power (e.g., the optical signal) and $G_X$ denotes $G_S$ or $G_B$ coming from the two different photoresponses described previously, we found that $G_S$ was the dominant gain (e.g., $G_S>10^4 G_B$) in these 2TDs at illumination levels typically used in photoconductance (PC) measurements, such as those reported in [3]. The title of this report, *"Study of defect states in GaN films by photoconductivity measurement"*, is wrong or at least greatly misleading because nobody measures the conductivity of materials, only the conductance $G=1/R$ of 2TDs containing these materials. To assume naively that conductance changes $\Delta G=1/\Delta R$ of 2TDs mirror changes in the conductivity of their inner materials *is wrong in general*. This assumption, which suggests that $G_B$ exists while $G_S$ is null, is false, and there is evidence of this. One piece of evidence can be found in [1, 2], where we showed that the most likely situation is the opposite one, where $G_S$ accounts well for the gain of the photoresistor because the $G_B$ contribution is negligible. More striking proof is found in [4], which shows that the enigmatic $1/f$ "excess noise" of solid-state devices is simply the noise mechanism that corresponds to the gain $G_S$ of these devices.

Given the low $P_L$ values used in PC measurements, the situation $G_S \gg G_B$ is very likely to occur. This indicates that the authors of [3] presuppose a gain mechanism for their GaN photoresistors that is likely wrong, although *their PC data* applied to the gain $G_S$ are excellent in explaining the yellow band (YB) found in the photoluminescence (PL) of n-GaN "samples", which are the Devices used in PL experiments because *nobody measures the PL of materials but the luminescence of devices* containing these materials. The devices required to measure PL do not need to have two terminals such as those 2TDs required in conductance measurements. In fact, they look like "simple" devices that only *require their surface* to be crossed by the laser light entering the material (n-GaN in this case) whose luminescent emission is then measured. A continuous medium such as an infinite volume of n-GaN material would be useless for PL measurements because it would not have the aforementioned surface separating the PL driver (laser) from the material driven (excited) by this laser that emits the luminescence we call the PL response or "PL spectrum" in general.

Therefore, *there is no way to avoid testing a device in a measurement system* because to extract information from the "device under test" (DUT) we need a system capable of exchanging the energy relevant to the experiment (e.g., photons in PL, electrical energy in conductance measurements, etc.). This demonstrates the essential role of the device in Physics, of which not all scientists are aware. When we consider this device and its effects on measurements, enigmatic measurements such "excess noise" in solid-state devices [4], flicker noise in vacuum devices [5] and even the more complex oscillator phase noise [6] are easily understood. Concerning the PL of n-GaN "material", its YB peak at approximately 2.2 eV [7] is enigmatic because it emits photons with energies below the bandgap $E_g$≈3.41 eV of this material at a temperature $T$=300 K (room $T$). To solve this "enigma", we must find a device capable of emitting photons with energies $hv<E_g$. Believing that the PL response is derived only from the bulk region, as most people do, we agreed that the solution presented in [7] (a handy trap in the volume of the n-GaN) was a good trial among the many fashionable theories on the subject, but one of us had *another proposal* for how this YB is generated, using the surface model used in [1, 2].

The GaN YB measured in [7] was assigned to transitions of electrons from the CB to a deep level lying at 1 eV from the VB. This "trap-based" model is a handy theory that is well accepted in journals; it is based on a bulk defect that we denote as being "ad hoc" in explaining the enigmatic effect. This leads to "materials doped by traps" that are difficult to understand because these traps are technologically "elusive". By this we mean that technology can provide two n-GaN materials with $N_d$ and with $5N_d$ donors per cm$^3$ quite reliably but not n-GaN with some YB emission and n-GaN with YB emission that is five times as strong. This suggests that the YB cannot be a property of the n-GaN material but the emission from something that exists in each PL arrangement we use to exchange optical energy with the n-GaN material. Because a device is the system we have to use for this exchange of energy, let us look for possible devices we may have overlooked in this case. This is *the other proposal* we had envisaged from a careful set of PC measurements we had performed at that time [1, 2]. However, its radical departure from the most fashionable theories on the YB of n-GaN in 1997 and its immaturity (it was based on [2], which was about to be published, and on [1], which had just been published), made it inappropriate for journals in which "differential changes" in a fashionable theory are likely accepted but radical departures are rejected due to the dogmatic defence of prevailing theories by most reviewers. Thus, the band diagrams shown in Fig. 1 did not appear in [7].

Using the results we obtained when we considered a physical model for a 2TD [8] that agrees with quantum physics [9], or better phrased, when we *do not forget the physics of the device used to extract information* (e.g., to measure), and being aware of the weak EL (electroluminescence) recently found in n-GaN field-effect transistors (FET) [10, 11], we decided to write this paper to communicate the *other proposal* we had in 1997 for the weak YB found in the PL of n-GaN samples. This radical proposal unifies the PC data of [3], the YB of the PL emission of n-GaN samples [7] and the weak EL of [10, 11] in a cogent model that explains these phenomena. Therefore, we consider the strong influence of the GaN surface in measurements, which we expected from the strong influence of the surface of GaAs devices [1, 2, 12]. The surface band-bending (SBB) of n-GaN surfaces, which is roughly twice that we measured in n-GaAs surfaces [2], is able to collapse GaN field-effect transistors (FETs) [13], hence the reason to expect its strong influence in PL experiments. Therefore, let us pay attention to the device linked to the surface of n-GaN material under photoluminescence (PL) conditions.

The key role of devices in measurements must be considered before assigning any property to their inner material. Refs. [1, 2, 4-6] demonstrate the benefits obtained when we do not assume naively that the conductance of a 2TD reflects the conductivity of its inner material. However, this is a common assumption that is reflected in the title of [3] and in the first sentence of [14], which cites only *two examples among the thousands* that appear in physics and engineering journals. As in GaAs Devices, this misconception about conductivity (in $\Omega^{-1}\text{cm}^{-1}$) mirroring (e.g., being proportional to) the conductance (in $\Omega^{-1}$) of photoresistors will also fail in GaN-based devices because their conductive volume is not equal to the whole volume of the GaN material. There are depleted regions under their surfaces because n-GaN tends to hold electrons in surface states (SEs). The SBB is derived from the double layer (DL) formed by the negative charges trapped in occupied SEs and the positive charges of a depleted region of thickness $W$. The collapse of GaN FETs is a dramatic effect of this behaviour of n-GaN surfaces that is capable of forming a negative floating gate on the surface of these devices [13, 15].

Although the manner in which the SBB invalidates the aforementioned proportionality was described in [1, 2], it was so deep-rooted in our minds that we wrote in the Introduction of [2]: "…*thus modulating the effective volume that takes part in the electrical conductivity of the samples*". It is clear that we meant "*in the electrical Conductance of the samples*", which are the 2TDs in which these frequency-resolved PC measurements were taken. Only from these conductance data and with some care can the conductivity of these materials be deduced. We apologise for making this mistake, which shows that we ourselves found it *very hard to abandon our prejudices* regarding conductivity modulation, which most people assume. Given the ease with surface devices in 2TDs are overlooked, this error is more likely to occur for devices without terminals such as the samples used in PL experiments. To demonstrate the misleading effects of this error, Section 2 will present a model for the surfaces of n-GaN samples that we will use to explain photoconductance (PC) measurements in n-GaN photoconductors. In Section 3, we will show how this model for the surface of n-GaN allows for the explanation of the YB of n-GaN samples under the optical driving to which they are submitted in PL experiments. Section 4 will demonstrate how this model allows for the prediction of the weak electroluminescence (EL) recently found in GaN-based transistors when holes reach their surfaces [10, 11], and some conclusions will be drawn at the end.

## 2. Surface-Induced Photoconductance in N-type GaN Devices

Though authors of [3] did not use an appropriate title for the report, their *photoconductance* data (PC data) are valuable in accurately modelling Device#1 due to the surface of n-GaN. Due to its depleted region under the surface, a GaN epilayer of thickness $d$ only offers a channel thickness ($d-W$) for the conductance of any 2TD using this epilayer as its channel between terminals. This was reported many years ago for GaAs [1, 2, 16]. Because a similar effect for the bottom interface between this epilayer and the substrate would roughly duplicate the reasoning at hand, it will not be dealt with at this moment. This bottom photo-FET reinforces our proposal regarding photoresistors presented in [3]: under a weak light source, the PC system will behave as a photo-FET, where photons will reduce the depleted thickness $W$ of their SBB, and not as photoresistors with noticeable conductivity modulation.

Thus, the name *photoconductive* detectors for the 2TD used in PC experiments is generally wrong because these photodetectors, which are made from two ohmic contacts on an epilayer, have two detection mechanisms $G_S$ and $G_B$ whose relative weight depends on the light power they receive (see Fig. 6 of [1] for details). Under high illumination power, they show the expected conductivity modulation by the photogenerated carriers giving them their photoconductive gain $G_B$, which is linked to the ratio between the photocarriers' lifetime and their transit time between the terminals of the 2TD. This is the type of response (gain $G_B$) assumed in [3], which requires a *hypothesis about the band tails of states* to allow photons with energies lower than $E_g$ to produce a PC response. At low illumination levels, however, these devices react as floating-gate photo-FETs whose associated SBB is reduced by photons. We attribute their upper surface (and bottom interface), whose DL looks like that of the gate of a FET, as being the source of their high gain $G_S$ due to the photobackgating of space-charge regions around the channel [1, 2]. This gain $G_S$, which is proportional to the inverse of the illumination power $P_L$ [1], *largely exceeds the gain $G_B$* assumed in [3], as we demonstrated in [1, 2] for GaAs with a similar but lower SBB.

Hence, under the low $P_L$ of a PC system, the photoresistors of [3] will exhibit a PC response because *$G_S$ does not require the generation of electron-hole pairs because the emptying of SEs is enough to produce a PC response* like that of GaAs [1, 2]. This PC response of the samples used in [3], which would not exist for photons with energy $hv<E_g$ without the band tails proposed in Fig. 3 for the gain $G_B$, is perfectly possible without band tails for the photo-FET gain $G_S$ by considering that photons with energy $hv<E_g$ empty occupied SEs without generating electron-hole pairs. This photo-induced SBB reduction and the shrinkage of $W$ it induces (light-induced backgating effect) will increase the conductance $G=1/R$ of the photoresistors described in [3], which under this weak illumination are *photoconductance* detectors of gain $G_S$ and not photoconductive detectors of gain $G_B$.

Hence, we propose that the PC signal exhibited by the devices described in [3] for photons with energy $hv<E_g$ is due to *photons emptying the SEs of their GaN surface*. To investigate this possibility, let us consider the band diagram shown in Fig. 1-a as we approach the surface of a sample of n-GaN in TE, thus in the dark. The flat part of Fig. 1-a is the band diagram of an infinite bulk n-GaN (Device#0) in TE. When Device#0 is "cut", the surface appears; thus, a density of $N_S$ cm$^{-2}$ SE is exposed on this surface, and *Device#1 is created* because $N_S$ allows for *new ways to store and exchange energy* by those electrons that occupy SEs in Device#1. This leads to eigenstates of energy such as those handled in the quantum treatment of this exchange called "Generalised Noise" in [9]. Concerning the charge states of these SEs linked to electrical energies, let us assume two choices derived from [13]: **a)** *neutral* when empty and **b)** *charged by a magnitude –q* C.

With electric charges arranged in space, we can use thermodynamics to say that Device#1 *has a degree of freedom to store electrical energy*. By this we mean that electrical energy is stored by the double layer (DL) or space-charge region formed by a sheet of negative charge in the surface due to $N_F$ cm$^{-2}$ electrons occupying SEs and a thicker sheet of positive charge (depleted region) of $+qN_F$ C/cm$^2$ compensating the $-qN_F$ C/cm$^2$ sheet density of the surface. This proximity of the $+qN_F$ and $-qN_F$ sheet charges minimises the stored energy, and its *dipolar nature* prevents the existence of electric fields far from the surface that would absorb energy for the conduction currents they would produce in the GaN bulk, for example. Therefore, this DL has $-qN_F$ C/cm$^2$ on the surface and a thicker slab of charge under the surface (a depleted region of thickness $W \approx N_F/N_d$ cm for a uniform doping $N_d$ cm$^{-3}$ of the n-GaN material). This DL creates the SBB shown in Fig. 1-a, which becomes an *energy barrier* $\Delta E = q\phi$ eV for those electrons in the conduction band (CB) of the n-GaN to reach the surface. This barrier selects those electrons of the CB liable to be trapped in SEs, thus constituting a *capture barrier*.

By viewing the surface as a *planar trap* with $(N_S - N_F)$ cm$^{-2}$ centres able to capture electrons from the GaN bulk, this capture process can be understood to be *thermally activated* [2] with energy $\Delta E = q\phi$ eV because only those electrons surpassing the SBB barrier can be captured by one SE (tunnelling is not considered to simplify the process). This *planar capture* differs from the capture assigned to a bulk trap with $N_T$ cm$^{-3}$ centres embedded in the volume because it is not an "in-situ" capture process that varies the *n* cm$^{-3}$ electron concentration in the CB. Instead, it is an "ex-situ" capture process in which each electron captured from the bulk GaN is not held within this bulk but at some distance on the surface. This distance, however, is crossed by some electrons of the CB, thus *allowing thermal interaction* by the exchange of particles and energy between the surface and the bulk. Hence, this capture by the surface *does not vary the concentration of electrons in the bulk* far from the surface as it would for a "bulk trap" with $N_T$ cm$^{-3}$ centres embedded in the GaN. This *planar capture slightly modifies the thickness W of the depleted region next to the surface* [2, 4].

In a planar channel with a thickness ($d$-$W$) under its surface, the fluctuations $\Delta W$ due to these emission-capture fluxes of electrons at the surface (see Fig. 1-a) will lead to *fluctuations $\Delta G$ in the conductance $G$* measured between the two terminals of the channel. These fluctuations $\Delta G$ and those emanating from fluctuations in the conductivity of its material, provided its thickness $W$ is constant, would be indistinguishable. Thus, assuming naively that $W$ is constant, we will find in the conductance of the channel of the 2TD the trapping effects due to this planar trap, which will mislead us towards a deep "bulk trap" with thermal energy $E_T=q\phi$ eV because emission-capture processes over a SBB are thermally activated with $E_T=q\phi$ eV [2]. This feature allowed us to measure SBBs by *photoconductance* frequency resolved spectroscopy [2], which previously was considered photoconductivity frequency resolved spectroscopy (PCFRS). This planar trap in the surface leads us to consider that the data of [3] are *excellent PC data that we will use* to determine the PC gain $G_S$ of these photoconductors operating as photo-FETs under low illumination (e.g., not conductivity data).

When photons with energy below $E_g$ reinforce the thermal emission indicated by an arrow in Fig. 1-a, the ($N_F/N_S$)$_{TE}$ ratio in TE (e.g., in the dark) is reduced to ($N_F/N_S$)$_{PC}$ by the weak light of the PC system. This is so because to sustain a higher (thermal+optical) emission in Device#1, we need more captures per second, which are only possible with a slightly lower SBB at the same temperature $T$. This lower SBB requires a lower $N_F$, which leads to a lower $W$, thus increasing the channel cross section of the photoresistor and hence its conductance. This gain $G_S$ gives the PC signals shown in [3], which the authors consider as being due to $G_B$, though without proof. Thus, the PC data for sample 511 that appear in Fig. 1 of [3] will have to do with a band diagram like that of Fig. 1-b but without holes in its valence band (VB) near the surface because *holes will not appear for photons with hv<$E_g$* in Device#1. Note that the SBB existing under illumination cannot be that in TE. If this were so, we would have ($N_F/N_S$)$_{TE}$=($N_F/N_S$)$_{PC}$, but this is not possible. If emission from the surface is higher because the electrons trapped in SEs have a lower lifetime $\tau$ under a higher thermal+optical emission, this requires more captures per unit time than in TE to be counterbalanced. This lowers $N_F$, which in turn reduces the SBB or capture barrier. Thus, the capture rate is slightly increased until a *dynamical equilibrium*, with $(N_F/N_S)_{PC}<(N_F/N_S)_{TE}$, is reached under the weak light of the PC system.

Our simple interpretation of the PC data of [3] as a photo-backgating effect due to the n-GaN SBB allows us to test our model of Fig. 1-a with the help of Fig. 2, which sketches the PC spectrum of sample 511 found in Fig. 1 of [3]. To complete our model for Device#1, let us assume that the SBB of n-GaN is $q\phi \approx 1.2$ eV (roughly twice the value we measured for n-GaAs surfaces [2]) and that the photovoltage $V_{ph}$ generated by photons with energy $h\upsilon<E_g$ of the PC system is small. In this case, Fig. 1-a would be mostly approximated for the sample under the weak light of the GE 1493 lamp filtered (thus attenuated) by a monochromator, as described in [3]. The absence of a PC signal for photons with $h\upsilon<1.3$ eV in Fig. 2 could mean that the light power with photon energy $h\upsilon<1.3$ eV was null in [3] or that the response of Device#1 (its gain $G_L$) for these photons was null. The first option is discarded because for p-doped GaN photoresistors such as sample AA1 of Fig. 1 of [3], a PC response was measured for $h\upsilon<1.3$ eV. Thus, the more likely situation is that *Device#1 did not produce a PC signal for photons with $h\upsilon<1.3$ eV*.

This would be so if the SBB of the n-GaN were close to 1.2 eV because those SEs above the Fermi level in Fig. 2 would be empty and would not produce any PC signal (e.g., no electrons could be emitted from empty SEs, and no change in $W$ would take place). This confirms the $q\phi \approx 1.2$ eV obtained for Device#1, but it does not indicate the absence of SEs for energies closer to the CB than $q\phi$: it only means that *SEs below the CB down to 1.2-1.3 eV are empty* at room temperature. With this model for Device#1 on the n-GaN surface, we could say that the PC system of [3] detects that sample 511 has occupied SEs up to ≈1.3 eV below the CB, and *from the slope* of its PC curve around point A in Fig. 2, we could say that *this distribution of SEs would continue to decrease as we approach the CB*. This behaviour, predicted from a trend (thus without data below the threshold energy $h\upsilon_t \approx 1.3$ eV), is only an attempt to predict what the SE density does near the CB. As the photon energy is increased from $h\upsilon_t \approx 1.3$ eV, the PC signal of sample 511 rises in a way that appears exponential from the semilog plot of Fig. 2. This suggests *an exponential increase in the density $N_S(E)$ as we go down in energy E towards the VB in the gap*, although we have to consider the cumulative effect of photo-emitted electrons as photon energy $h\upsilon$ increases. By this we mean that the PC photon flux with energy $hυ≈2$ eV not only empties SEs lying at 2 eV below the CB (see the horizontal arrow in Fig. 2) but also those SEs lying between 2 eV and 1.3 eV below the CB (see arrows indicating all of these contributions for photon energy $hυ=2$ eV in Fig. 2).

For each photon energy $hυ$, the PC response or gain $G_L$ *accumulates effects due to all of the occupied SEs* from the uppermost SE close to the Fermi level $E_F$ in Fig. 1-a (properly speaking, close to the surface imref $E_{Fs}$ of Fig. 1-b) to the lowest SE lying $hυ$ eV below the CB that photons are able to empty. Because these types of calculations were published many years ago by one of the authors of [3] in an excellent book [17] (see its Section 3-A-5), we can confidently say that the band tails proposed in Fig. 3 of [3] simply contribute $N_S(E)$, which is *the energy distribution of SEs within the gap of n-GaN*. From the PC data of [3], we obtain an $N_S(E)$ proportional to $\exp(hυ/E_0)$ with, $E_0$ between 230 meV and 280 meV [3]. Thus, the PC curve of sample 511 sketched in Fig. 2 shows that *there is a band of SEs from $hυ≈1.3$ eV below the CB (at least) to the VB itself*, whose $N_S(E)$ increases as $\exp(hυ/E_0)$ as we approach the VB. The density of bulk states shown in Fig. 3-a of [3] would reflect this $N_S(E)$ of SEs, not the tails of states its authors believe are indicated by the deep-rooted "conductivity modulation" concept. To further prove this result, we will find in Section 4 *empirical proof of the validity of this $N_S(E)$ derived from the PC data of [3] and our model for the surface of n-GaN.*

Concerning the *tails of states close to the VB* proposed in [3], it is worth noting that the abrupt increase in PC data for $hυ≈E_g$ (close to 3.41 eV at room $T$) in Fig. 1 of [2] (see Fig. 2) may have to do with these types of tails. In this case, however, the response of the 2TD of [3] would start to show its response $G_B$ in the bulk in addition to the response $G_S$ linked to the surface we have considered up to this point. Thus, the title of [3] would not be as misleading because it might appear from the beginning of this section that its sample 511 has *two devices generating its PC signal*: Device#1, due to its surface, producing the main response for $hυ<E_g$ at this low illumination power and Device#0, due to its bulk region, adding a PC response more closely linked to $G_B$ for $hυ>E_g$, both contributing to the PC signal measured between their two terminals. In any case, we can conclude that on the surface of n-GaN there is a band of SEs going from the VB up to ≈1.3 eV below the CB (and most likely up to the CB itself) with a density $N_S(E)\propto\exp(-hυ/E_0)$. This band explains the great ability of n-GaN to trap electrons at SEs, which produces its high SBB and its harmful effects in GaN FETs [13, 15].

To show that the surface photovoltage $v_{ph}$ under the weak illumination of the PC system is negligible, let us consider Fig. 1-a, where the planar capture by the surface and the planar emission from the surface in TE are mutually counterbalanced to maintain the average SBB of n-GaN devices. We use "average" because this SBB is derived from a dynamical equilibrium in which the *fluctuation in W it endures* is an undesired source of conductance noise [4, 12]. Therefore, let us consider this *planar source of conductance fluctuations* in channels such as those used by most solid-state devices today. Because the fluctuations in the conductance of a 2TD due to a planar trap and those expected for a bulk trap are indistinguishable, planar traps strongly suggest the existence of deep levels lying hundredths of meV below the CB. This notion is reinforced by the high thermal activation energy $E_T = q\phi$ eV the traps show, which is what we believed to be true before discovering that emission-capture processes over a SBB show a thermal activation energy $E_T = q\phi$ eV in PC experiments [2]. Thus, the planar traps on the surface perfectly mimic the effects expected for the handy carrier traps known as "deep levels". This imitation of a non-existent bulk trap goes beyond the appearance of a high thermal activation energy $E_T = q\phi$ eV in Arrhenius plots as soon as $T$ varies by a few tens of K. *This imitation is so good that* the electron flux captured by the surface ($CA$ in cm$^{-2}$/s) becomes

$$CA = c_n \times (N_S - N_F) \times n \times \exp\left(\frac{-q\phi}{kT}\right) \quad (1)$$

where $k$ (J/K) is the Boltzmann constant, $n$ (cm$^{-3}$) is the free electron gas concentration in the GaN bulk under the surface, ($N_S$-$N_F$) (cm$^{-2}$) is the density of SEs able to capture electrons from the bulk and $C_n$ is a capture coefficient, which must be expressed in cm$^3$/s, *the same units of the capture coefficient $C_n$ of the "well-known" bulk traps*. This means that $C_n$ can be taken as the product of the mean thermal velocity $v_{th}$ (cm/s) of electrons moving towards the surface and a capture cross section $\sigma_n$ (cm$^2$), which, when combined with the exponential term of the SBB barrier, leads to the idea of *a thermally activated capture cross section for an non-existent deep trap in the bulk*.

Therefore, let us gain deeper insight into this planar trap by considering the emission flux counterbalancing the capture flux expressed by Eq. (1) in TE. Using the concept of lifetime $\tau$ for the $N_F$ cm$^{-2}$ electrons occupying SEs, whose thermal activity will emit from time to time towards the GaN bulk, the emitted flux ($EM$ in cm$^{-2}$/s) will be

$$EM = \frac{N_F}{\tau} = e_n N_F \quad (2)$$

Eq. (2) indicates that a pure emission transient from the GaN surface towards the GaN bulk without the counteracting capture that it uses to exist when trapping transients in devices will experience an exponential decay of $N_F$ with lifetime $\tau$. However, the boundary conditions during these transients lead to *processes in which both emission and capture coexist* in time *t*. A good example of such species are donor-related transients in AlGaAs, where a net-capture transient starting without electrons in donors at *t*=0 (thus with null Emission at *t*=0) undergoes continually enhanced emission as the number of trapped electrons in donors (DX centres) increases with time [18]. This means that a net capture process will have a negative feedback (NF) or "shutter" due to **i)** increasing emission as the number of donors able to emit electrons to the CB increases with *t* and **ii)** decreasing capture as the number of donors able to capture electrons from the CB decreases with *t*. The transient thus produced ends when the rising emission counterbalances the decreasing capture, the whole process being expressed by Eq. (3) of [18]. This is a Riccati equation, where capture is proportional to the square of the free electron concentration $n^2(t)$ because the concentrations of free electrons $n(t)$ and empty donors are equal under the *boundary condition of charge neutrality*, preventing the electron gas from escaping during capture. In searching for the counterpart of this Riccati equation for the emission-capture processes, maintaining the average charge density $qN_F$ C/cm$^2$ on the surface of GaN, we have

$$\frac{\partial N_F}{\partial t} = CA - EM \approx c_n \times N_S \times n \times \exp\left(\frac{-q\phi}{kT}\right) - \frac{N_F(t)}{\tau} \quad (3)$$

Eq. (3) has been simplified for clarity by considering a SE density that is much higher than the occupied SE density ($N_S \gg N_F$). Although we will not solve this non-linear equation here, we will say that the emission-capture process it represents features an *NF mechanism* through which it is difficult to terminate its transients because, added to the increasing emission and decreasing capture during the transient, each electron captured at instant *t* increases the capture barrier, thus greatly reducing the ability of the system to make further captures. By this we mean that electrons captured at instant *m* do not *linearly* reduce the capture at instant (*m*+1) as reported in [18]. Capture at instant (*m*+1) is reduced *exponentially*, thus producing a more brusque termination of the emission-capture transient than that under the conditions described in [18]. We could say that the NF or shutter ending the transients of DX centres is "soft", whereas the shutter ending the surface-related transients in Fig. 1 is "very brusque".

In the log-lin plot shown in Fig. 3, whose slope shows the time departure of the aforementioned net-capture transients from pure exponential decays (straight lines in this plot), the "set of slopes" (instantaneous lifetimes) for Eq. (3) solved for a SBB $q\phi\approx1.2$ eV is wider than those shown in Fig. 2 of [18]. This has been sketched in Fig. 3 to illustrate why surface-related transients in GaN devices such as HFETs have to be fitted by stretched exponentials [15]. This would be a *typical signature of these planar traps*, where charge neutrality must be maintained for their DL during the transient and *the capture barrier is modulated* by trapped electrons as the surface $N_F(t)$ increases with $t$. This leads to transients that recall the left side of the bathtub curve widely used in reliability engineering (see Figs. 8 and 9 of [15] for example) when they are viewed with linear axis.

Equating Eq. (3) to zero means that the emission flux is counterbalanced by the capture flux *on average* because the exact balance at each instant is impossible when the emission and capture processes are uncorrelated at this level. Therefore, the reduction in $\tau$ due to optical emission reinforcing thermal emission will reduce $N_F$, thus leading to a slightly lower capture barrier (e.g., a thinner DL), and the conductance of two-terminal devices using the GaN epilayer as their conducting channel will rise as we have assumed. This can be verified by reducing $\tau$ to $\alpha\tau$ ($\alpha<1$). If we maintain the density $N_F$ of occupied SEs in TE, the higher EM that results will require a lower capture barrier under illumination to be counterbalanced and thus a lower $N_F$ as we have assumed. Therefore, the number of occupied SEs $(N_F)_{PC}$ of the n-GaN device under the weak illumination of the PC system is $(N_F/N_S)_{PC}<(N_F/N_S)_{TE}$, which requires a lower depleted thickness $W$, which produces the PC signal. The low photovoltage $V_{ph}$ we have assumed in Fig. 1-a as a good approximation of Fig. 1-b in PC experiments is derived from the fact that $V_{ph}\approx18$ mV at room temperature would be enough to counterbalance twice the thermal emission at room temperature, which is hard to believe under the weak light of the PC system. Therefore, we can assume confidently that $V_{ph}<<\phi$ for the SBB $q\phi\approx1.2$ eV, which our model uses to explain the PC data of [3] and the YB of n-GaN in the next Section.

## 3. Surface-Assisted Luminescence in N-type GaN Devices

The photoluminescence (PL) spectrum of n-GaN samples shows a rather weak yellow band (YB) centred approximately 2.2 eV, such as that shown in Fig. 3-a of [7], where this YB was explained as being due to transitions of electrons from the CB to a deep level lying at 1 eV from the VB. To explain *the other proposal* we had for this YB, Fig. 1-b shows the band diagram of Device#1 and Device#0 existing in an n-GaN sample used in PL experiments under the strong illumination of the PL laser. As it is quite well known, photons creating electron-hole pairs in GaN reduce its SBB [1, 2] (see also Sec. 18-B-2 of [17]). This barrier reduction, which will increase the flux of electrons from the n-GaN bulk towards the surface, is due to a smaller depletion thickness $W$ that has a lower positive charge under the surface. Due to the dipolar nature of the DL, this requires a reduction in the negative charge at the surface, thus suggesting a lower occupation of the $N_S$ SE at the surface than in TE: $(N_F/N_S)_{PL} < (N_F/N_S)_{TE}$.

This lower occupation of the SE contradicts the higher capture flux that appears when the SBB is reduced. This higher capture should produce the opposite result: the SE *should be more occupied* than in TE because the thermal capture over this weakened capture barrier (e.g., over the lower SBB) simply means that the rate of electrons spilling over SE is higher. "Spilling" reflects the "electron fall over empty SE" under these conditions, where, beyond the capture barrier, the high flux of electrons arriving at the surface can be trapped by any SE, including those between $E_{Fs}$ and the bottom of the CB, which were "empty" under TE and under the weak illumination of the PC system for photons with energy $h\nu < E_g$. They are also empty in the PC system because the weak light power handled in PC experiments is likely unable to induce a noticeable "*pumping action*" over the thermal activity, which makes the imref $E_{Fs}$ a kind of sharp borderline that separates the occupied SEs below $E_{Fs}$ from the empty SEs above $E_{Fs}$; however, this will change under the strong illumination of the PL laser, as will be discussed.

Going back to the increased capture under PL conditions, let us consider how a *higher occupation of SEs*, as this capture suggests $(N_F/N_S)_{PL} > (N_F/N_S)_{TE}$, *with a lower negative charge density at the surface* is required to reduce the capture barrier or to flatten the SBB to some extent. The key element to achieve this in Device#1 is the *screening effect of hole*s swept towards the surface by the electric field of the SBB. Taking $N_H$ cm$^{-2}$ as the sheet density of holes accumulated at the surface, the negative sheet charge at the surface becomes $-(qN_F)_{PL} + (qN_H)$. In this way, the trapped charge at SEs can be higher than in TE if there are enough accumulated holes. This solves the electrostatic problem, making a reduction in SBB with a higher $N_F$ possible, but it opens a new way to reduce $N_F$, which is through the possible recombination of these electrons trapped in SEs with those holes accumulated in the surface as well.

This recombination, or in better terms its radiative part observed in PL, is the *other proposal* we formulated in 1997 to explain the origin of the weak YB observed in [7], which we will now justify. From Fig. 3 of [7], where this YB was magnified by 500 times, we can say that this recombination is very weak; otherwise, such yellow emission would be strong enough to make efficient yellow emitters from n-GaN material. We will later discuss the reason for this weakness. Therefore, we can assume that holes accumulated at the surface of n-GaN do not noticeably affect the lifetime of electrons trapped in SEs under PL conditions such that the increased capture would lead to the aforementioned $(N_F/N_S)_{PL}>(N_F/N_S)_{TE}$ condition by the screening effect of $N_H$. This *accumulation of holes at the surface* allows for the lower SBB shown in Fig. 1-b due to some photovoltage $V_{ph}$ that reduces the capture barrier to $q(\phi-V_{ph})$ eV. In this way, the emission and capture fluxes mutually counterbalanced in a steady state during the PL experiment at the surface are higher than those in TE.

It is worth noting that the $N_F$ electrons/cm$^2$ trapped at the surface can be higher than those in TE as explained or lower than those in TE because this depends on $N_H$ and therefore on the laser illumination power. What matters is to realise that we have $N_F$ and $N_H$ in close proximity at the surface, and the next point to consider is the effect of the optical emission of the PL laser emptying SEs lying below the imref $E_{Fs}$ in Fig. 1-b. This has to do with the "*pumping action*" mentioned previously in this section. Under the strong light of the PL laser, we cannot assume that the pumping action is negligible with respect to the thermal activity, which we assumed under the weak light of the PC system. Under PL conditions, we could expect a *noticeable pumping of electrons* trapped in SEs under the imref $E_{Fs}$, which would become a less sharp borderline than in TE between occupied SEs below $E_{Fs}$ and empty SEs above $E_{Fs}$. Instead, we should expect a broader transition band in energy, likely centred at $E_{Fs}$, separating the occupied SEs well below $E_{Fs}$ from empty SEs well above $E_{Fs}$ due to the aforementioned spill-over of electrons being captured over the new barrier (e.g., the SBB lowered by the PL laser light).

The strong illumination of the PL laser not only suggests a local heating of the sample but also the possibility of a high efficiency of the PL laser to empty SEs to achieve a flat band condition by greatly reducing $N_F$. If this were so, the SBB would disappear without the help of holes at the surface. In this case, the bands would be flat during the PL experiment and we would only obtain the PL spectrum of GaN from to band-to-band transitions with energy close to the bandgap $E_g$≈3.41 eV at room temperature, and no YB would be found in this case. Under this hypothetical flat-band condition, the capture barrier would be null and a huge flux of electrons would reach the surface from the GaN bulk to occupy SEs efficiently emptied by the laser pumping action. This is hard to believe, however, because photons can increase the energy of electrons trapped in SEs (e.g., vertical transitions in energy), but they are *unable to transport the negative charge* of all of these electrons accumulated at the surface.

The negative charge of electrons thus accumulated at the surface would produce a SBB sweeping them towards the GaN bulk, while other electrons from the bulk would be trapped over this SBB by SEs to be pumped subsequently by the laser. This would generate a SBB that contradicts the flat-band condition we have used as a starting point of this reasoning and will not be considered further, although a careful study of the power ratio $P_{YB}/P_{PL}$ between the power of the yellow band $P_{YB}$ and that of the PL spectrum expected for the n-GaN material could provide additional information about all of these processes related to Device#1 at the surface of n-GaN. Concerning the flattening of the SBB, we can say that $V_{ph}$ *will be low* because the hole charge $N_H$ accumulated at the surface cannot be increased without limit because as it increases, the depth of the well for holes at the surface diminishes, allowing holes to escape. If $V_{ph}$≈18 mV at room temperature, the net capture and net emission would be *doubled*. Assuming that $(N_F/N_S)_{PL}$≈$(N_F/N_S)_{TE}$, this doubled emission would derive from $(N_F)_{PL}$≈$2(N_F)_{TE}$ (see Eq. (2)), which in turn would require $N_H$≈$(N_F)_{TE}$ to keep a SBB close to that in TE. This $N_H$ value starts to fill the well shown in Fig. 1-b for holes. In other words, the $N_H$ charge able to accumulate at the surface *is limited*, and this fact together with the need for an appreciable SBB to transport electrons from the surface towards the bulk (recall that photons can transport them) limits $V_{ph}$ during the PL experiment.

It is worth noting that the band diagrams of Fig. 1 apply to our undoped wurzite GaN layers of [7], whose *residual doping was n-type* in the $10^{17}$-$10^{18}$ cm$^{-3}$ range. This means that the ratio $(N_F/N_S)_{PL}$ of [7] during the PL experiments was most likely higher than in TE or that the SEs of the n-GaN during our PL experiments were likely more occupied than in TE due to photogenerated holes swept towards the surface, as shown in Fig. 1-b. Device#1 features *many electrons in SEs and many holes in close proximity*, which leads us to consider the *emission with carrier interaction* described in Sec. 6-D-2-e of [17], where a photon with energy $hv<E_g$ was produced when an electron of the CB made "a transition to a *virtual state a* at an energy $\Delta E$ below the CB by exciting an electron inside the CB to a higher-energy state (this causes a change in momentum). The first electron completes the transition from state *a* to the VB by emitting a photon $hv$".

Again from [17], we "note that momentum-conservation rules make this process difficult to observe in pure direct-bandgap semiconductors since an additional phonon may be needed.... then the transition becomes a three-step process having a very low probability of occurrence". In Fig. 1-b, however, we do not need a third electron promoted to a higher state of energy to have electrons in virtual states (VSs) because *electrons trapped in SEs can reach VSs close to the surface by tunnelling* through thin barriers whose heights are not far from the SBB (see Fig. 1-b). Hence, the small fraction of $(N_F)_{PL}$ electrons reaching VSs in a region abundant with holes would complete the transition to the VB by *emitting photons with energy $hv<E_g$*.

These photons would peak at an energy $E_{peak}\approx 2.2$ eV; this value is derived from the difference $E_g-q\phi\approx 2.21$ eV that appears by using $q\phi\approx 1.2$ eV for the n-GaN we obtained from the PC data of [2] in Section 2; moreover, the quasi-Fermi level of the SE system ($E_{Fs}$) will not vary much with respect to its value in TE ($E_F$) because $V_{ph}$ is low, as explained previously, and $N_S$ is much higher than $N_F$. This high $N_S$ in GaN seems likely because if it was low, no relevant negative floating gate would appear in FETs [13] and no passivation of their surfaces would be required. The YB for energies higher than 2.2 eV would be due to those SEs filled above $E_{Fs}$ in combination with the "pumping action" of the PL laser emptying some SEs below $E_{Fs}$ and the filling of SEs above $E_{Fs}$ by the enhanced capture. Those electrons in SEs above $E_{Fs}$ would tunnel to VS with a higher probability than electrons in SEs below $E_{Fs}$, which are much more numerous, however, due to the exponential dependence of $N_S(E)$. From the product of this occupation density of SEs and the probability of electrons in these SEs of tunnelling to VS, a peak energy not far from $E_{Fs}$ would result, thus giving rise to a peak in the YB around $E_{peak}\approx 2.2$ eV.

This would be the origin of the YB peak observed in the PL spectrum of n-GaN in Fig. 3 of [7], whose *oscillating character* around $E_{peak}$ could reveal an oscillating behaviour of the tunnelling probability of these electrons with respect to their corresponding virtual states or an interference-related phenomenon requiring further study. In any case, this surface-assisted luminescence (SAL) is the plausible reason we have formulated for the YB peaking at ≈2.2 eV in the PL of n-GaN samples. Because this SAL is due to electrons trapped in SEs like those forming the negative floating gate of n-GaN FETs in the dark [13], it allows for the prediction of surface electroluminescence (SEL) when holes are injected near the surface of GaN devices, as has been observed recently in n-GaN FETs [10, 11]. In these studies, the SEL has been attributed to "hot-electron effects" due to its decay with photon energy, which closely tracks the $N_S(E)$ that we obtained in Section 2 from the PC results of [3], as will be discussed in the next section.

## 4. Surface Electroluminescence in N-type GaN Devices and Other Effects

The surface-related origin of the SAL known to explain the YB of GaN not only agrees with many studies linking this YB to the surface [19-23] but also allows for the prediction that if holes are brought near the surface in n-GaN devices, *surface electroluminescence (SEL) proportional to the surface charge $N_F$ will be observed.* Let us show that this SEL is very likely the weak EL found recently in the FET devices of [10-11], where it has been attributed to hot-electron effects. Using our model for Device#1 at the surface, we do not need hot-electron effects at all, in agreement with recent studies discarding these types of effects in the electrical degradation of GaN FETs [24]. Our model also predicts that *this SEL will track the dynamics of the surface charge $N_F$* in n-GaN devices and that *this SEL will not be a yellow peak but a "redder" luminescence.*

Recalling the role of Device#1 in the formation of the negative floating gate (NFG) that collapses these FETs [13], we can explain why *this SEL is first observed in the gate side towards the drain* of field-effect devices for moderate drain voltages $V_{DS}$ [10] and why it *moves towards the drain contact* if the voltage $V_{DS}$ is increased further [11]. Due to the positive feedback (PF) that underlies this electrical degradation and the subsequent collapse of GaN FETs [13], this SEL will appear near the drain only *after some delay* and after *surpassing a threshold* voltage $V_{DS}$ representing the onset of the electrical degradation. This delay has been reported in [11], and the threshold was reported in [24]. These two features—a *delay* to allow the PF to build the collapse *and a threshold* to trigger this degradation—leading to collapse are familiar features of circuits with PF, such as the one proposed in [13] ten years ago for the electrical degradation collapsing GaN FETs. Thus, the time evolution of this SEL for large $V_{DS}$ [11] gives strong support to our model based on surface Device#1, which applies to different phenomena such as the PC response of [3], the YB of [7] and the SEL of [10-11].

Regarding its "colour", this SEL will not peak like the YB because there is no "pumping action" of a laser in [10-11] as it occurs in the PL when the YB appears. The FET-like devices of [10-11] *were placed in the dark* to observe their weak EL, which only appears when holes are injected near the GaN surface. Therefore, the spectrum of their SEL *will reflect the exponential increase in $N_S(E)$ as we go from the CB to the VB* in Fig. 4, which is the counterpart of Fig. 2 for surface Device#1 acting as a photon emitter and not as the photon absorber it was in Fig. 2. Hence, this SEL will not peak at $E_{peak} \approx 2.2$ eV as the YB of the PL does. Instead, it will show an exponential decay in intensity as the energy of the emitted photons increases. To be precise, this SEL has to show an intensity proportional to $\exp(-h\nu/E_0)$; thus, it will possess the same slope α shown in Fig. 2, though decreasing with photon energy $h\nu$ because the density $N_S(E)$ in the GaN surface decreases as $\exp(-h\nu/E_0)$ when we go from the VB to the CB. This $N_S(E) \propto \exp(-h\nu/E_0)$ was obtained in Section 2 from the PC data of [3] obtained 17 years ago, and this SEL observed recently in [10, 11] *is the empirical proof we promised in Section 2 of its validity*, which is ensured when we consider that $E_0$ was between 230 meV and 280 meV [3]. Fig. 4 shows a sketch of the weak EL spectrum shown in Fig. 3-b of [10].

For $E_0$=230 meV, we have $E_0 \approx 2600$ K which means that the slope of Fig. 4 will be tg(α)=-1/$E_0$, as *if this SEL was the EL derived from a hot-electron gas at $T_{eq} \approx 2600$ K*. This is the explanation provided in [10] for the decay of the weak EL the authors observed with this $T_{eq}$. Unaware of the Device#1 existing at n-GaN surfaces, these authors propose that the weak EL they observe is derived from a hot-electron gas in the GaN channel at this $T_{eq} \approx 2600$ K, an effect that our model does not require at all. This $T_{eq}$ in [10] confirms that the PC data of [3], with a proper model for Device#1, reveal the density of surface states at the surface of n-GaN devices, as we have proposed. This is good proof of the validity of the relationship $N_S(E) \propto \exp(-h\nu/E_0)$ derived using our model from the PC data obtained long ago [3], though it has not been well understood until now. Although the time evolution of this SEL moving from the gate contact to drain one after some delay from the application of a sufficiently high $V_{DS}$ [11] provides additional proof about the origin and usefulness of this SEL as a marker for surface charge in n-GaN devices, we will consider this point further due to the space constraint of this paper.

## Conclusions

The physical relevance of the devices used in measurements must be considered before assigning properties to bulk materials. All hypotheses used to assign the result of a measurement of a device to a property of a bulk material must be verified. This forces researchers to consider the devices associated with the surfaces and interfaces of actual materials, and only after having taken their effects into account can we consider their finite volume (Device#0) as a good model for a continuous medium or infinite volume of these materials.

The yellow band assigned to GaN is better understood if we consider it as the surface-assisted luminescence of n-GaN due to the unintentional Device#1 existing at its surface. Device#1 also helps to understand the photoconductance response of 2TDs made from n-GaN materials, especially at low illumination levels. Finally, the same surface Device#1 allows for the prediction and thus the explanation of the weak electroluminescence found in n-GaN devices when holes are injected near their surfaces. Device#1 at the surface of GaN provides a unified view of the luminescence phenomena of this material, which is difficult to obtain using other models that are unaware of this surface device.

## Acknowledgements

This work is supported by the Spanish CICYT under the MAT2010-18933 project, by the Comunidad Autónoma de Madrid through its IV-PRICIT Program and by the E. U. project Nº 304814 RAPTADIAG.

# Figure captions

**Fig. 1. a)** Band diagram of n-GaN Device#1 near its surface in thermal equilibrium (TE). **b)** Band diagram of Device#1 out of TE during a photoluminescence experiment.

**Fig. 2.** Diagram linking photoconductance signal (log axis) to photon energy for Device#1, whose band diagrams are shown in Fig. 1. The PC signal at point A is due to electrons emitted from SE close to the quasi-Fermi level $E_F$, while the PC signal at point B accumulates the emissions of electrons trapped at SEs covering a band between $E_F$ and 2 eV below the CB (see the text).

**Fig. 3.** Semilog plots of net-capture transients derived from traps with the same emission lifetime $\tau$ but different boundary conditions during the transient. During the surface-like decay, the capture barrier increases as capture takes place (see text).

**Fig. 4.** Semilog plot of the surface luminescence spectrum that must appear in n-GaN devices when holes are injected near their surface. Electrons in surface states (SE) using virtual states to accomplish radiative transitions to the valence band (see text) would reflect the exponential distribution with the energy of the SEs.

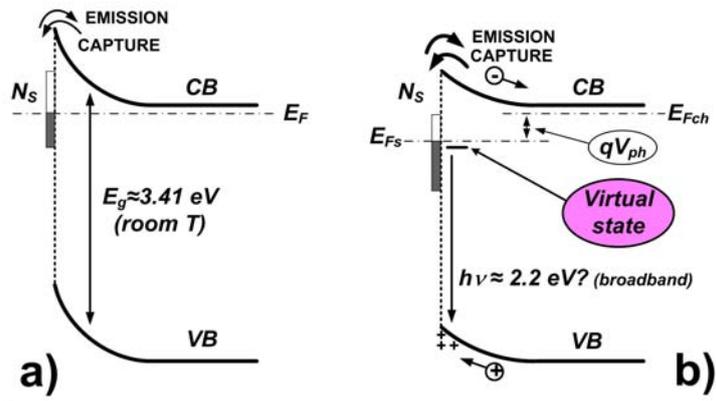

**Figure 1**

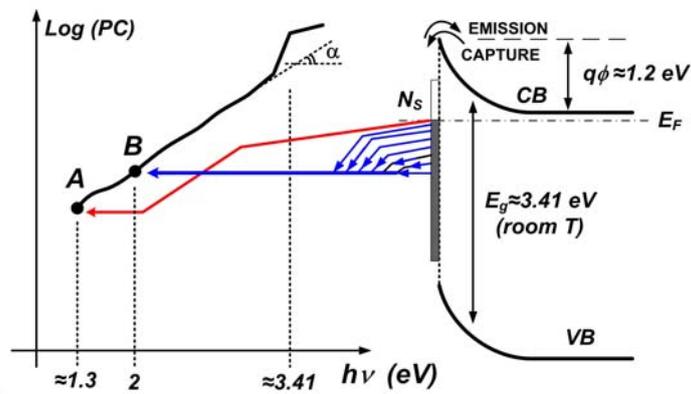

**Figure 2**

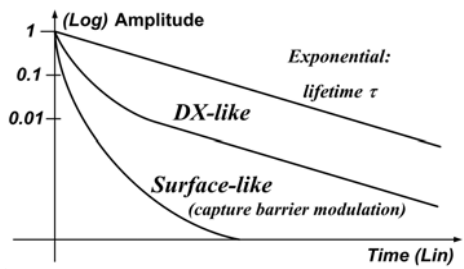

**Figure 3**

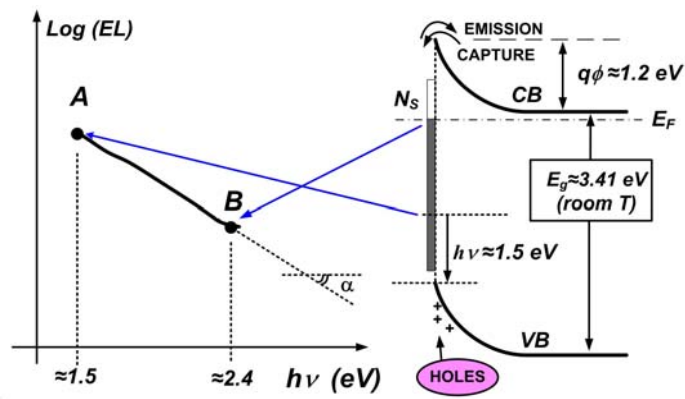

**Figure 4**
23